\documentstyle[preprint,aps]{revtex}
\tightenlines
\begin{document}
\draft
\title{A quantum-geometrical description of the statistical laws of nature}

\author{Wellington da Cruz}
\address{Departamento de F\'{\i}sica,\\
 Universidade Estadual de Londrina, Caixa Postal 6001,\\
Cep 86051-970 Londrina, PR, Brazil\\
E-mail address: wdacruz@exatas.uel.br}
\date{\today}
\maketitle
\begin{abstract}

We consider the fractal characteristic of the quantum mechanical paths 
and we obtain for any universal class of fractons labeled by 
the Hausdorff dimension defined within the interval 
$1$$\;$$ < $$\;$$h$$\;$$ <$$\;$$ 2$, a fractal distribution function 
associated with a fractal von Neumann entropy. Fractons are 
charge-flux systems defined in two-dimensional multiply connected space 
and they carry rational or irrational values of spin.

This formulation can be considered in the context of the fractional 
quantum Hall effect-FQHE, where we discovered that the quantization of 
the Hall resistance occurs in pairs of dual topological quantum numbers, the 
filling factors. In this way, these quantum numbers get their topological 
character from the Hausdorff dimension associated with the 
fractal quantum path of such particles termed fractons. On the other hand, 
the universality classes of the quantum Hall transitions can be classified 
in terms of $h$. Another consequence of our approach, which is supported by 
symmetry principles, is the prediction of the FQHE. The connection between Physics and 
Number Theory appears naturally in this context.
\\
keywords: Fractal distribution function; fractal von Neumann entropy; 
Fractons; fractional quantum Hall effect; number theory.
\end{abstract}
\newpage

We make out a review of some concepts introduced by us in the literature, 
such as\cite{R1,R2,R3,R4,R5,R6,R7,R8}: fractons, universal classes $h$ of particles, 
fractal spectrum, duality symmetry betwenn classes $h$ of particles, 
fractal supersymmetry, fractal distribution function, 
fractal von Neumann entropy, fractal index etc. We apply these 
ideas in the context of the FQHE and number theory. 

Fractons are charge-flux systems which carry rational or irrational 
values of spin. These objects are defined in 
two-dimensional multiply connected space and are classified in 
universal classes $h$ of particles or quasiparticles, with the 
fractal parameter or Hausdorff dimension $h$ , defined in the interval 
$1$$\;$$ < $$\;$$h$$\;$$ <$$\;$$ 2$. It is related to the 
quantum paths and can be extracted from the propagators 
of the particles in the momentum space\cite{R2,R9}. The particles are collected in 
each class take into account the fractal spectrum

\begin{eqnarray}
&&h-1=1-\nu,\;\;\;\; 0 < \nu < 1;\;\;\;\;\;\;\;\;
 h-1=\nu-1,\;
\;\;\;\;\;\; 1 <\nu < 2;\nonumber\\
&&h-1=3-\nu,\;\;\;\; 2 < \nu < 3;\;\;\;\;\;\;\;\;
 h-1=\nu-3,\;
\;\;\;\;\;\; 3 <\nu < 4;etc.
\end{eqnarray}

\noindent and the spin-statistics relation $\nu=2s$, 
valid for such fractons. The fractal spectrum establishes a 
connection between $h$ and the spin $s$ of the particles: 
$h=2-2s$, $0\leq s\leq \frac{1}{2}$. Thus, there exists a mirror 
symmetry behind this notion of fractal spectrum. 
Given the statistical weight for these classes of fractons

\begin{equation}
\label{e11}
{\cal W}[h,n]=\frac{\left[G+(nG-1)(h-1)\right]!}{[nG]!
\left[G+(nG-1)(h-1)-nG\right]!}
\end{equation}

and from the condition of the entropy be a maximum, we obtain 
the fractal distribution function\cite{R2}

\begin{eqnarray}
\label{e.44} 
n[h]=\frac{1}{{\cal{Y}}[\xi]-h}
\end{eqnarray}

The function ${\cal{Y}}[\xi]$ satisfies the equation 

\begin{eqnarray}
\label{e.4} 
\xi=\biggl\{{\cal{Y}}[\xi]-1\biggr\}^{h-1}
\biggl\{{\cal{Y}}[\xi]-2\biggr\}^{2-h},
\end{eqnarray}

\noindent with $\xi=\exp\left\{(\epsilon-\mu)/KT\right\}$. We understand the 
fractal distribution function as a quantum-geometrical 
description of the statistical laws of nature, 
since the quantum path is a fractal curve and this 
reflects the Heisenberg uncertainty principle. 

We can obtain for any class its distribution function considering 
Eq.(\ref{e.44}) and Eq.(\ref{e.4}). For example, 
the universal class $h=\frac{3}{2}$ with distinct values of spin 
$\biggl\{\frac{1}{4},\frac{3}{4},\frac{5}{4},\cdots\biggr\}_{h=\frac{3}{2}}$, 
has a specific fractal distribution

\begin{eqnarray}
n\left[\frac{3}{2}\right]=\frac{1}{\sqrt{\frac{1}{4}+\xi^2}}.
\end{eqnarray}

\noindent We also have
 
\begin{eqnarray}
\xi^{-1}=\biggl\{\Theta[{\cal{Y}}]\biggr\}^{h-2}-
\biggl\{\Theta[{\cal{Y}}]\biggr\}^{h-1}
\end{eqnarray}

\noindent where

\begin{eqnarray}
\Theta[{\cal{Y}}]=
\frac{{\cal{Y}}[\xi]-2}{{\cal{Y}}[\xi]-1}
\end{eqnarray}

\noindent is the single-particle partition function. 
We verify that the classes $h$ satisfy a duality symmetry defined by 
${\tilde{h}}=3-h$. So, fermions and bosons come as dual particles. 
As a consequence, we extract a fractal 
supersymmetry which defines pairs of particles $\left(s,s+\frac{1}{2}\right)$. 
In this way, the fractal distribution function appears as 
a natural generalization of the fermionic and bosonic 
distributions for particles with braiding properties. Therefore, 
our approach is a unified formulation 
in terms of the statistics which each universal class of 
particles satisfies: from a unique expression 
we can take out any distribution function. In some sense , we can say that 
fermions are fractons of the class $h=1$ and  
bosons are fractons of the class $h=2$.

The free energy for particles in a given quantum state is expressed as

\begin{eqnarray}
{\cal{F}}[h]=KT\ln\Theta[{\cal{Y}}].
\end{eqnarray}

\noindent Hence, we find the average occupation number

\begin{eqnarray}
\label{e.h} 
n[h]&=&\xi\frac{\partial}{\partial{\xi}}\ln\Theta[{\cal{Y}}].
\end{eqnarray}

\noindent The fractal von Neumann entropy per state in terms of the 
average occupation number is given as\cite{R1,R2} 

\begin{eqnarray}
\label{e5}
{\cal{S}}_{G}[h,n]&=& K\left[\left[1+(h-1)n\right]\ln\left\{\frac{1+(h-1)n}{n}\right\}
-\left[1+(h-2)n\right]\ln\left\{\frac{1+(h-2)n}{n}\right\}\right]
\end{eqnarray}

\noindent and it is associated with the fractal distribution function Eq.\ref{e.44}.

The entropies for fermions $\biggl\{\frac{1}{2},
\frac{3}{2},\frac{5}{2},\cdots\biggr\}_{h=1}$
 and bosons $\biggl\{0,1,2,\cdots\biggr\}_{h=2}$, can be recovered promptly
 
\begin{eqnarray}
{\cal{S}}_{G}[1]=-K\biggl\{n\ln n +(1-n)\ln (1-n)\biggr\} 
\end{eqnarray}

\noindent and
 
\begin{eqnarray}
{\cal{S}}_{G}[2]=K\biggl\{(1+n)\ln (1+n)-n\ln n\biggr\}. 
\end{eqnarray}

\noindent Now, as we can check, each universal class $h$ of particles, 
within the interval of definition has its entropy defined 
by the Eq.(\ref{e5}). Thus, for fractons of the self-dual class
$\biggl\{\frac{1}{4},
\frac{3}{4},\frac{5}{4},\cdots\biggr\}_{h=\frac{3}{2}}$, we have
  
\begin{eqnarray}
{\cal{S}}_{G}\left[\frac{3}{2}\right]=K\left\{(2+n)\ln\sqrt{\frac{2+n}{2n}}
-(2-n)\ln\sqrt{\frac{2-n}{2n}}\right\}. 
\end{eqnarray}

\noindent We have also introduced the topological concept of fractal index, 
which is associated with each class. As we saw, $h$ is a geometrical parameter 
related to the quantum paths of the particles and so, we define\cite{R3} 

\begin{equation}
\label{e.1}
i_{f}[h]=\frac{6}{\pi^2}\int_{\infty(T=0)}^{1(T=\infty)}
\frac{d\xi}{\xi}\ln\left\{\Theta[\cal{Y}(\xi)]\right\}.
\end{equation}

\noindent We obtain for the bosonic class $i_{f}[2]=1$, 
for the fermionic class $i_{f}[1]=0.5$ 
and for some classes of fractons, we have 
$i_{f}[\frac{3}{2}]=0.6$, $i_{f}[\frac{4}{3}]=0.56$, $i_{f}[\frac{5}{3}]=0.656$. 
For the interval of the definition $ 1$$\;$$ \leq $$\;$$h$$\;$$ \leq $$\;$$ 2$, there 
exists the correspondence $0.5$$\;$$ 
\leq $$\;$$i_{f}[h]$$\;$$ \leq $$\;$$ 1$, which signalizes 
the connection between fractons and quasiparticles of the conformal field theories, 
in accordance with the unitary $c$$\;$$ <$$\;$$ 1$ 
representations of the central charge. For $\nu$ even it is defined by 

\begin{eqnarray}
\label{e.11}
c[\nu]=i_{f}[h,\nu]-i_{f}\left[h,\frac{1}{\nu}\right]
\end{eqnarray}

\noindent and for $\nu$ odd it is defined by 

\begin{eqnarray}
\label{e.12}
c[\nu]=2\times i_{f}[h,\nu]-i_{f}\left[h,\frac{1}{\nu}\right],
\end{eqnarray}

\noindent where $i_{f}[h,\nu]$ means the fractal 
index of the universal class $h$ which contains the particles 
with distinct values of spin, which obey specific 
fractal distribution function. For example, we obtain the results

\begin{eqnarray}
&&c[0]=i_{f}[2,0]-i_{f}[h,\infty]=1;\nonumber\\
&&c[1]=2\times i_{f}[1,1]-i_{f}[1,1]=0.5;etc.
\end{eqnarray}

\noindent In another way, the central charge $c[\nu]$ can be obtained using the 
Rogers dilogarithm function, i.e. 

\begin{equation}
\label{e.16}
c[\nu]=\frac{L[x^{\nu}]}{L[1]},
\end{equation}

\noindent with $x^{\nu}=1-x$,$\;$ $\nu=0,1,2,3,etc.$ and 

\begin{equation}
L[x]=-\frac{1}{2}\int_{0}^{x}\left\{\frac{\ln(1-y)}{y}
+\frac{\ln y}{1-y}\right\}dy,\; 0 < x < 1.
\end{equation}

\noindent Thus, we have established a connection between fractal geometry and 
number theory, given that the dilogarithm function appears 
in this context, besides another branches of mathematics\cite{R10}. 

Such ideas can be applied in the context of the FQHE. This phenomenon is 
characterized  by the filling factor parameter $f$, and for 
each value of $f$ we have the 
quantization of Hall resistance and a superconducting state 
along the longitudinal direction of a planar system of electrons, which are
manifested by semiconductor doped materials, i.e. heterojunctions, 
under intense perpendicular magnetic fields and lower 
temperatures\cite{R11}. 

The parameter $f$ is defined by $f=N\frac{\phi_{0}}{\phi}$, where 
$N$ is the electron number, 
$\phi_{0}$ is the quantum unit of flux and
$\phi$ is the flux of the external magnetic field throughout the sample. 
The spin-statistics relation is given by 
$\nu=2s=2\frac{\phi\prime}{\phi_{0}}$, where 
$\phi\prime$  is the flux associated with the charge-flux 
system which defines the fracton $(h,\nu)$. According to our approach 
there is a correspondence between $f$ and $\nu$, numerically $f=\nu$. 
This way, we verify that the filling factors observed 
experimentally appear into the classes $h$ and from the definition of duality 
between the equivalence classes, we note that the FQHE occurs in pairs 
 of these dual topological quantum numbers\\

 $(f,\tilde{f})=\left(\frac{1}{3},\frac{2}{3}\right), 
\left(\frac{5}{3},\frac{4}{3}\right), \left(\frac{1}{5},\frac{4}{5}\right), 
\left(\frac{2}{7},\frac{5}{7}\right),\left(\frac{2}{9},\frac{7}{9}\right), 
\left(\frac{2}{5},\frac{3}{5}\right), \left(\frac{3}{7},\frac{4}{7}\right), 
\left(\frac{4}{9},\frac{5}{9}\right) etc$.\\

We verify that all the experimental results satisfy this symmetry principle.

Finally, we observe again that our formulation to 
the universal class $h$ of particles with any values of spin $s$ 
establishes a connection between Hausdorff dimension $h$ and 
the central charge $c[\nu]$. Besides this, we have obtained a relation between the 
fractal parameter and the Rogers dilogarithm function, through the 
concept of fractal index, which is defined 
in terms of the partition function associated with each universal class of particles. 
Also we have established a connection between the fractal 
parameter $h$ and the Farey 
sequences of rational numbers. Farey series $F_{n}$ of order 
$n$ is the increasing sequence of 
irreducible fractions in the range $0-1$ whose 
denominators do not exceed $n$. We have the following 

{\bf Theorem}\cite{R6}: {\it The elements of the Farey series $F_{n}$ 
of the order $n$, belong to the fractal sets, whose Hausdorff 
dimensions are the second fractions of the fractal sets. The 
Hausdorff dimension has values within the interval 
$1$$\;$$ < $$\;$$h$$\;$$ <$$\;$$ 2$, which are associated with fractal curves.}

For a extended review about our work see Ref.\cite{R8}.


\begin{thebibliography}{99}
\bibitem{R1} W. da Cruz, Physica {\bf A313} (2002), 446.
\bibitem{R2} W. da Cruz, Int. J. Mod. Phys. {\bf A15} (2000), 3805.
\bibitem{R3} W. da Cruz and R. de Oliveira, Mod. 
Phys. Lett. {\bf A15} (2000), 1931.
\bibitem{R4} W. da Cruz, J. Phys: Cond. Matter. {\bf 12} (2000), L673.
\bibitem{R5} W. da Cruz, Mod. 
Phys. Lett. {\bf A14} (1999), 1933.
\bibitem{R6} W. da Cruz, Chaos, 
Solitons and Fractals,  {\bf 17} (2003), 975.
\bibitem{R7} W. da Cruz,  cond-mai/0301587.
{\it On the dual topological quantum numbers filling factors.}
\bibitem{R8} W. da Cruz, Int. J. Mod. Phys. {\bf A18} (2003), 2213. {\it 
Proceedings 2nd International Londrina Winter School: 
Mathematical Methods in Physics, August, 26-30, 2002}. 
Eds. M. C. B. Abdalla {\it et al.} 
\bibitem{R9} A. M. Polyakov, in {\it Proc. Les
 Houches Summer School
 {\bf vol. IL}}, ed. E. Br\'ezin and J. Zinn-Justin
  (North Holland, 1990) 305.
\bibitem{R10} A. Kirillov, Prog. Theor. Phys. Suppl. {\bf 118} (1995), 61.
\bibitem{R11} R. B. Laughlin, Rev. Mod. Phys. {\bf 71}, (1999), 863;\\
H. Stormer, Rev. Mod. Phys. {\bf 71}, (1999), 875;\\
D. C. Tsui, Rev. Mod. Phys. {\bf 71}, (1999), 891;\\ 
and references therein.
\end{thebibliography}
\end{document}